# Statistical Validation in Cultural Adaptations of Cognitive Tests: A Multi- Regional Systematic Review

1st Miit Daga
School of Computer Science Engineering and Information Systems
Vellore Institute of Technology
Vellore, India
miitdaga03@gmail.com

2nd Priyasha Mohanty
School of Computer Science Engineering and Information Systems
Vellore Institute of Technology
Vellore, India
priyashamohanty03@gmail.com

3rd Ram Krishna
School of Computer Science Engineering and Information Systems
Vellore Institute of Technology
Vellore, India
ramkrishna50505@gmail.com

4th Swarna Priya RM
School of Computer Science Engineering and Information Systems
Vellore Institute of Technology
Vellore, India
swarmapriya.rm@vit.ac.in

*Abstract*— This systematic review discusses the methodological approaches and statistical confirmations of cross-cultural adaptations of cognitive evaluation tools used with different populations. The review considers six seminal studies on the methodology of cultural adaptation in Europe, Asia, Africa, and South America. The results indicate that proper adaptations need holistic models with demographic changes, and education explained as much as 26.76% of the variance in MoCA-H scores. Cultural-linguistic factors explained 6.89% of the variance in European adaptations of MoCA-H; however, another study on adapted MMSE and BCSB among Brazilian Indigenous populations reported excellent diagnostic performance, with a sensitivity of 94.4% and specificity of 99.2%. There was 78.5% inter-rater agreement on the evaluation of cultural adaptation using the Manchester Translation Evaluation Checklist. A paramount message of the paper is that community feedback is necessary for culturally appropriate preparation, standardized translation protocols also must be included, along with robust statistical validation methodologies for developing cognitive assessment instruments. This review supplies evidence-based frameworks for the further adaptation of cognitive assessments in increasingly diverse global health settings.

## I. Introduction

Cognitive impairment and dementia are two of the most significant threats to public health globally. Dementia is a neurodegenerative disorder with progressive nature, which means it deteriorates multiple cognitive domains. These include memory, thought processes, decision- making, spatial awareness, ability to learn, understanding, and language skills [9]. Mild Cognitive Impairment (MCI) refers to an intermediate stage that falls between normative cognitive functioning and dementia, thus affecting multiple aspects of cognitive abilities of the elderly population [9].

According to World Health Organization, globally, there are 47.5 million people who have dementia, and 58% of these are from LMICs. Each year, 7.7 million new cases of dementia occur, and the total is predicted to double by 2030 and nearly triple by 2050 [12]. Likewise, the world population is also ageing rapidly and the population of the elderly, those aged 60 years and above, is considered to be doubled by 2050 [13].

Moreover, studies on dementia continue to underrepresent Indigenous populations, which restricts our knowledge of how the illness affects different groups and what factors contribute to its prevalence [10][11]. Thus, the need for tools that are culturally appropriate for the assessment of cognition is quickly becoming critical across the globe, especially in relation to healthcare systems dealing with more diverse populations. Testing instruments conventionally designed primarily within Western paradigms often ignore cultural nuances, differences in education, and language usage that can dramatically impact the results of the tests and the accuracy of conclusions from diagnostic tests. This has deep and far-reaching implications: patients are being wrongly diagnosed with cognitive disabilities due to assessment tools inappropriate to their culture when they are still cognitively intact; conversely, people that are actually suffering from cognitive damage are being overlooked due to screening tools that ignore cultural differences. Other than simple translation, there are other challenges posed in changing these cognitive assessment tools [7]. The barriers include educational inequality, cultural perceptions of cognitive functioning, and varying literacy rates within different populations. For instance, tasks that depend on the understanding of Western beliefs or require specific experiences based on education may systematically place certain populations at a disadvantage and lead to false assumptions about cognitive impairments. Previous studies have documented cases where culturally inappropriate measures have led to high misclassification rates as high as which points out the need for proper adaptation of the measures. This literature review aims to discuss and synthesize existing methodological approaches related to cross-cultural adaptation of cognitive assessment measures, focusing on:

1. Statistical validation methods and their generalizability across different cultural contexts.
2. How demography influences the test and adaptation requirements.
3. Protocol of cultural adaptation and its effects on accuracy.
4. Translation and adaptation process standardization across different populations.

This systematic review of adaptation methodologies aims to provide evidence-based frameworks for healthcare practitioners and researchers involved in cross-cultural cognitive evaluation. By understanding effective adaptation strategies and their statistical validation, it is possible to enhance the precision of cognitive assessments across varied global populations, thereby ultimately elevating diagnostic accuracy and patient care outcomes.

## II. Literature Review

The reviewed studies consistently highlight the importance of cultural sensitivity in adapting cognitive assessment tools. Ogbuagu et al. (2024) demonstrated a rigorous adaptation process in Southeast Nigeria, adhering to International Test Commission (ITC) guidelines, and underscored the necessity of cultural relevance in cognitive testing. Similarly, Mirza et al. (2023) introduced the Manchester Translation Evaluation Checklist (MTEC), which evaluated cultural nuances in the Mini-Mental State Examination (MMSE) Urdu version with 78.5% inter-rater agreement [3][6].

Nielsen (2022) outlined the adaptations across European countries and highlighted the challenges posed by linguistic barriers, literacy levels, and cultural contexts for minority groups such as South Asians in the UK and Turkish populations in Germany. These challenges were elaborated by Theocharous et al. (2024), which focused on a study of the Montreal Cognitive Assessment for hearing-impaired individuals (transa-H) across four European languages, showing a 6.89% variance of scores due to cultural-linguistic factors. This was necessary as MoCA is largely administered through speech and up to 70% of adults aged over 60 years have hearing impairment which can cause misdiagnosis or overestimation of cognitive impairment [1][2][8].

## III. Material and Methods

### A. Participants and Demographics

The studies reveal a wide-ranging cross section of respondents. For instance, the sample in Bezerra et al.'s (2024) study in Brazil comprised 141 indigenous participants from Brazil. They belonged to 34 ethnic groups, with a mean age of 61.9 years (SD±9.2) and education ranging between 0 to 18 years, with 15.6% illiterate. Similarly, Nepal et al.'s (2019) study reflected that 60% were uneducated; therefore low-literacy-specific adaptations will be necessary for such groups [4][5].

In Europe, Theocharous et al. (2024) reported on participants with a mean age of 74.51 years (SD±8.77). It also noted significant variation across language groups, and the mean age of French participants was reported to be 82.39 years. Gender was balanced, and 55.32% of participants had education less than 12 years. Ogbuagu et al.'s (2024) study in Nigeria reported the gender representation to be equally distributed and the educational backgrounds of the healthcare workers and representatives from the community [2][3].

The demographic diversity across studies (Table 1) underscores the importance of considering educational and age-related factors in cognitive assessment adaptations.

### B. Test Battery Selection and Statistical Analysis

The cognitive assessment tools varied among the studies, but they all focused on critical cognitive domains. Bezerra et al. (2024) used the Mini-Mental State Examination (MMSE), Brief Cognitive Screening Battery (BCSB), Verbal Fluency tests, and the Pfeffer Functional Activities Questionnaire, achieving significant diagnostic accuracy with sensitivity and specificity of 94.4% and 99.2%, respectively. On the other hand, Ogbuagu et al. (2024) used subtests from the Brain Health Assessment (BHA), focusing on memory, language skills, executive functioning, motor abilities, and visuospatial skills [3][5].

The MoCA-H study by Theocharous et al. (2024) revealed cultural-linguistic performance differences of up to 2.6 points, with a need for demographic adjustments. Age, sex, and education explained a total of 26.76% of the variability in MoCA-H scores [2].

### C. Language Adaptation Protocol

The reviewed studies used the robust translation and adaptation procedures. Nepal et al. (2019), used a four-step process: forward translation, synthesis, review by the expert committee, and pre-testing. The technical terms such as "visuospatial orientation" were discussed in the consensus to overcome the difficulty of translation. Nielsen (2022) set up protocols for native language assessments and standard interpreter services, while the study by Theocharous et al. (2024) showed significant variations in performance despite standardized translations, indicating the need for culture-specific adaptations [1][2].

### D. Educational Level Adjustment and Cultural Framework

Education was an influential factor on performance in the cognitive assessments. Ogbuagu et al. (2024) from Nigeria attempted to counteract educational disparity by adopting 'visuoconstructional' tests. It modified the cube-drawing component by using stick designs, where it proved a 89% rate and reduced failure. Theocharous et al. (2024) validated the use of a 2-point adjustment for those who had ⩽12 years of education [2][3].

Nielsen's (2022) ECLECTIC framework ensured that the assessment tool was contextually appropriate for education, cultural factors, language proficiency, and communication styles. Similarly, the study by Mirza et al. (2023) ensured that the modification was relevant to the culture by introducing visual and content adjustments from community feedback [6].

### E. Statistical Validation and Quality Assurance

Validation procedures in the studies used stringent metrics. The content validity index scores of the study by Mirza et al. (2023) (S-CVI = 0.93 for equivalence aspects) also showed that its adapted instruments were reliable. Internal consistency (Cronbach's alpha = 0.7) and structured focus groups complemented validation in the study by Nepal et al. (2019) [4][6].

TABLE I. PARTICIPANT DEMOGRAPHICS ACROSS STUDIES

| Study ID | Region | Sample Size | Age Range (Mean ± SD) | Gender (% Female) | Education Level (% Illiterate) |
|---|---|---|---|---|---|
| 2 | Europe | 385 | 74.51 ± 8.77 | 46.23 | 55.32% with ≤12 years |
| 3 | Nigeria | 30 | 64.0 ± 15.9 (HCW), 64.1 ± 11.9 (Community) | 53.30 (HCW) | Varied (9.5 ±0.4 years HCW) |
| 4 | Nepal | 100 | 67.72 ± 6.08 | 46.00 | 60% |
| 6 | Brazil | 141 | 50-90 (61.9 ± 9.2) | 58.20 | 15.60% |

a. HCW = Healthcare Workers; SD = Standard Deviation. Studies 1 and 5 did not report complete demographic information in the comparable format and are thus excluded from this demographic summary. Education levels are reported in different formats across studies based on their primary measurement approach.

### F. Cultural Modification Process

The modifications towards the cultural factors involved languages, pictures, and activity-related adaptations. In iterative cycles of the study by Mirza et al. (2023), photo-based pictures were replaced by line drawings, culturally acceptable for such representation and incorporating known symbolic expressions. Bezerra et al.'s (2024) MTEC-based evaluations under the experiment pin pointed actual issues that require adaptation regarding the culture, for instance, an English-based task to spell "WORLD" backward. Table 2 summarizes the various cultural adaptation methods employed across the reviewed studies, highlighting the diverse approaches taken to ensure cultural equivalence and validity [5][6].

TABLE II. SUMMARY OF CULTURAL ADAPTATION METHODS ACROSS STUDIES

| Study ID | Cultural Adaptation Methods | Key Features |
|---|---|---|
| 3 | ITC Guidelines, trilingual competency (English, Igbo, Pidgin) | Emphasized cross-cultural equivalence and test validity |
| 4 | Four-stage translation process | Included forward translation, synthesis and expert review. |
| 5 | MTEC Evaluation Framework | Focused on semantic, technical, content, conceptual, and criterion equivalence |
| 6 | Community feedback and iterative refinement | Integrated culturally appropriate visual elements and task modifications |

b. ITC = International Test Commission; MTEC = Manchester Translation Evaluation Checklist. Studies 1 and 2 focused on European-wide adaptations but did not specify singular cultural adaptation methods and thus are not included in this summary table.

## IV. RESULTS AND DISCUSSION

### A. Results

Major conclusions from the literature review include adaptation and validation of cognitive assessment tools not only in different cultures and demographics but also towards specific cultural and demographic considerations. These conclude that without cultural sensitivity, adjustments due to demographic factors, and strong statistical methodology in developing adapted tools, results could be meaningless.

*1) Effectiveness of Cultural Adaptation*
- The Manchester Translation Evaluation Checklist has been shown in this instance of cultural adaptation to have reliability in terms of inter-rater agreement at 78.5% regarding the Urdu translation of Mini-Mental State Examination; this can only be obtained with a focus on semantic, conceptual, and cultural equivalence.
- Iterative approaches, such as in the case of the study by Mirza et al. (2023), found that involving community stakeholders had a significant impact on increasing the cultural relevance of the adapted tools. For example, replacing photographs with culturally appropriate line drawings improved participants' understanding and participation [6].

*2) Impact of Demographics on Performance*
- Education levels consistently impacted the performance on cognitive assessments. Studies by Theocharous et al. (2024) and Nepal et al. (2019), emphasized the necessity for demographic modifications, including particular adjustments such as score recalibrations for individuals with 12 years or fewer of education [2][4].

- There was evidence for language-specific differences in the MoCA-H study; the cultural-linguistic factors accounted for 6.89% of the variance in scores. The maximum difference in performance was about 2.6 points between English, French, Greek, and German groups, pointing to the role of linguistic adaptations.

*3) Statistical Validation*
- Diagnostic accuracy was very high with the adapted tools. For example, the sensitivity and specificity of the MMSE from the study by Mirza et al. (2023) were 94.4% and 99.2%, respectively. In addition, content validity indices for the tools gave their reliability in S-CVI = 0.93 for equivalence [6].

- The use of statistical metrics like Cronbach's alpha ($\geq 0.7$) ensured the internal consistency of adapted instruments, providing a benchmark for future studies.

Table 3 presents the comprehensive statistical results and validation metrics across studies, demonstrating the robustness of the adapted assessment tools through various statistical measures.

### B. Discussion

Findings from these reviewed studies present a rich understanding of challenges and solutions in adapting cognitive assessment tools for diverse populations. Many critical themes arise from such a discussion:

*1) Integration of Cultural Sensitivity and Community Involvement*
- Cultural adaptation encompasses more than mere linguistic translation. Research has consistently highlighted the necessity of considering cultural subtleties, including the significance of test content and the incorporation of symbols or tasks that are culturally recognizable. For instance, Bezerra et al.'s (2024) incorporation of feedback from Indigenous communities in Brazil resulted in tools that were more congruent with cultural contexts, thereby improving diagnostic precision and participant acceptance. [5]

- The iterative refinement process, as demonstrated in the research by Nepal et al. (2019) and Mirza et al. (2023), emphasizes community engagement. This will ensure that modified tools will be relevant to local communities and effectively overcome both linguistic and cultural obstacles [4][6].

TABLE III. STATISTICAL RESULTS AND VALIDATION METRICS ACROSS STUDIES

| Study ID | Assessment Tool | Statistical Metrics | Cultural-Linguistic Impact | Validation Results |
|---|---|---|---|---|
| 2 | MoCA-H | - 26.76% score variance explained by demographics (age, sex, education)<br>- 6.89% variance due to cultural-linguistic factors | 2.6-point performance difference across language groups | Score adjustment of +2 points validated for ≤12 years education |
| 4 | Nepal Adaptation | Cronbach's α ≥ 0.7<br>Four-stage validation process | Translation challenges identified in technical terminology | Internal consistency validated through structured focus groups |
| 5 | MMSE (Urdu) | 78.5% inter-rater agreement<br>Content validity assessed through MTEC | Identified significant adaptations needed for English-centric tasks | Cultural equivalence validated through systematic review process |
| 6 | MMSE & BCSB | 94.4% sensitivity<br>99.2% specificity<br>S-CVI = 0.93 for equivalence | Line drawings preferred over photographs for cultural relevance | High diagnostic accuracy achieved post-cultural adaptation |

c. Note: MoCA-H = Montreal Cognitive Assessment for hearing-impaired; MMSE = Mini-Mental State Examination; BCSB = Brief Cognitive Screening Battery; MTEC = Manchester Translation Evaluation Checklist; S-CVI = Scale-level Content Validity Index

*2) Demographic Adjustments as a Central Pillar*

- Education is a key determinant that affects performance in test taking. Replacing cube drawing exercises with stick drawings in the third study of low-literacy Nigeria groups offered a pragmatic solution for the resolution of educational inequities. These changes aid in the maintenance of cognitive tests that are both accessible and precise for different literacy levels.

- Age-related and gender-related variations were also controlled using specific methodologies that ensured that demographic variables did not affect the diagnostic findings.

*3) Methodological Rigor and Statistical Validation*

- Use of organized frameworks such as MTEC and ECLECTIC ensured robust methodologies that would provide cultural and linguistic equivalence. Such frameworks offer replicable procedures for subsequent adaptations and highlight technical, semantic, and conceptual consistency.

- The sensitivity, specificity, and internal consistency measures had to be high in the statistical validation of adapted tools to ascertain their validity. High accuracy rates, like MMSE's sensitivity at 94.4%, further add importance to strict validation procedures.

- The high statistical validation metrics presented in Table 3, including sensitivity rates of 94.4% and specificity of 99.2%, demonstrate the effectiveness of these cultural adaptations.

*4) Ethical Considerations and Consent Processes*

- Ethical compliance was an integral component of the studies, with explicit institutional approvals and participant consent. In this regard, the study by Nepal et al. (2019) documented ethical committee approvals and informed consent processes, ensuring transparency in the study and adherence to ethical standards. [4]

Future studies should continue to prioritize ethical considerations, particularly when working with vulnerable populations, to uphold research integrity and participant trust.

V. CONCLUSION

The elaborate review further discusses the complicated issues and innovative strategies required to adapt cognitive evaluation instruments for use in demographically and culturally diverse populations. This adaptation is beyond mere technical modifications and represents a very crucial step toward making healthcare services inclusive and culturally sensitive.

The main findings show that tight methodological frameworks such as MTEC and ECLECTIC need to be integrated to systematically create pathways toward cultural and linguistic equivalence. Community feedback and stakeholder involvement have been found to be important in ensuring the relevance and acceptability of these tools for various populations. Participatory approaches tend to foster a deeper understanding of local contexts, thus enabling changes that are not only more than linguistic translation but also better able to address cultural and educational inequalities.

The significance of demographic modifications is paramount. By altering 'visuoconstructional' tasks for populations with low literacy and considering variations in performance specific to language, these investigations emphasize the critical need for customizing cognitive evaluations to align with the distinctive attributes of the intended populations. Evaluation metrics such as sensitivity, specificity, and internal consistency substantiate the effectiveness of these adjustments, reinforcing the imperative for empirical rigor within this field.

Ethical considerations are also critical, as they ensure that adaptations respect the rights of participants and ensure that research integrity is preserved. Proper documentation of ethical clearances and consent procedures in most studies ensures a standard that subsequent research endeavors follow.

In planning ahead, the establishment of standardized methodologies for adaptation—yet to allow room for changes suitable to local specificities—will be critical. Intensification of interdisciplinary cooperation between linguists, clinicians, statisticians, and cultural specialists is likely to further advance the development of cognitive assessment tools that are both robust and applicable across populations. Technological advances and other developments in digital forms of cognitive assessments may open new avenues to address problems with scalability and accessibility.

In conclusion, the results and discourse put forward in this study provide an imperative for scholars, policymakers, and practitioners in the healthcare sector to highlight cultural competence in the assessment of cognitive health. This way, it may pave for equitable healthcare outcomes by focusing on all populations in this global effort to improve cognitive health.